# Do the technical universities exhibit distinct behaviour in global university rankings? A Times Higher Education (THE) case study

**Carmen Perez-Esparrells[a*] and Enrique Orduna-Malea[b]**

[a] Universidad Autónoma de Madrid, Einstein, 3, 28049, Madrid, Spain.
[2] Universitat Politècnica de València, Camino de Vera s/n, 46022, Valencia, Spain.
* Corresponding author: carmen.perez@uam.es

**Abstract**: Technical Universities (TUs) exhibit a distinct ranking performance in comparison with other universities. In this paper we identify 137 TUs included in the THE Ranking (2017 edition) and analyse their scores statistically. The results highlight the existence of clusters of TUs showing a general high performance in the Industry Income category and, in many cases, a low performance on Research and Teaching. Finally, the global score weights were simulated, creating several scenarios that confirmed that the majority of TUs (except those with a world-class status) would increase their final scores if industrial income was accounted for at the levels parametrised.

## 1. Introduction

In the current academic globalised scenario, top-tier universities continue to develop all the three university missions (teaching, research, knowledge transfer) but with different mission mixes. However, since the emergence of global university rankings, as well as the obsession with achieving world-class status (Cheng, Wang and Liu, 2014; Douglass, 2016), higher education institutions (HEIs) have focused disproportionately on research (Hazelkorn, 2015; Margison, 2017). This is because the methodology of these global rankings are biased towards research performance, the main driver of ranking outcomes, and about reputation (Margison, 2017), without distinguishing the numerous functions universities must fulfil, especially teaching and knowledge transfer (Laredo, 2007; Montesinos et al., 2008; Sarrico et al., 2010).

The reason for focusing on research rankings is that “research is an international activity and reasonable indicators exist for comparing institutions. Education, by contrast, is largely organised nationally and reflects different cultures and traditions” (Butler, 2010). However, even if one considers only the research-oriented methodologies, there is disagreement on which data, methodology and interpretations are more accurate. While Olcay and Bulu (2017) highlight significant differences among the leading global rankings (including THE, ARWU, QS, Leiden and URAP), even when measuring the same research criteria, others advocate for the need to use multi-faceted data (Moed, 2017) and clusters (Johnes, 2018), since there is no composite that reflects all the dimensions that shape a university. In line with this, Frenken, Heimeriks and Hoekman (2017) demonstrate how research performance differences among universities stem mainly from their sizes, disciplinary orientations and locations. Moreover, the bibliometric method that is used to rank research institutions fails to capture the full range of university research productivity (Van Raan, 2005) and selected criteria, and indicators are usually flawed due to the available data (Orduna-Malea, 2012; Gomez-Sancho and Perez-Esparrells, 2012).

Consequently, global rankings have caused an unintended effect of pressure towards homogeneity in order to achieve world-class status (Altbach and Salmi 2011), because the "second mission" rankings calculated through knowledge metrics (journal metrics, article-level metrics, author-level metrics) prevail in the new framework of international higher education, where the flow of knowledge between academics and industry is weak (Ribeiro-Soriano and Berbegal-Mirabent, 2017). This might explain the lack of "third mission" rankings. Moreover, those who are interested in second mission rankings are closer to the academic community than to the rest of society (Montesinos et al. 2008).

Apart from the general shortcomings regarding the design and operation of global rankings, which has been widely discussed in the literature (Harvey, 2008; Bookstein et al., 2010; Rauhvargers, 2011), these league tables are likely to obscure the profile of specific institutions, such as the technical universities (TUs), which make up approximately 12% of the universities worldwide (Frenken, Heimericks and Hoekman, 2017).

Due to the fact that the globalisation of Higher Education, Science and Technology, HEIs has become more central, complex and increasingly influential (Albatch, 2007) and TUs are called upon to play a leading role (De la Torre and Perez-Esparrells, 2017). However, global rankings do not reflect certain outcomes related to technology transfer, innovation and commercialisation, because they overlook some important features of the TUs (for instance the production of patents, the dissemination of technical results, the establishment of spinoffs and start-ups, etc.), thus masking the merits of institutions with a more entrepreneurial and technical profile. The authors of this study call for greater attention to be paid to how global rankings bodies measure the performance of TUs and the side effects produced by the indicators and weights, taking into account this case study of the THE ranking.

Some global rankings bodies attempt to characterise the purposes for which universities exist, such as knowledge transfer. This is the case with the THE, which attempts to capture commercialisation activities and advisory work in an innovative approximation through Industry Income scores. However, this is a simplistic perspective because the transfer of university technology to industry involves a multitude of mechanisms, which can be broken down into an even larger number of activities (see Hsu et al. 2015).

In the case of TUs, the interaction between the university and the activities developed in the scientific and technological ecosystem (technology transfer offices, science parks that promote research and innovation, employment bureaus, and entrepreneur programmes and incubators) are generally of greater intensity than in comprehensive universities. The level of university-business cooperation and innovative partnership with the knowledge economy is much higher as well (Minguillo and Thelwall, 2015). According to Caldera and Debande (2010), the agglomeration of knowledge close to universities has a positive effect on universities' technology transfer performances. This fact, together with the difficulty in producing indicators on knowledge transfer and innovation at international scale, entail hard work for the ranker editors so as to generate a new form of global rankings for knowledge transfer and innovation assessment, or a new typology of world-class TU ranking systems.

Therefore, a new approach to rankings would be to consider world-class universities (WCUs) from a non-traditional perspective. Despite the fact that a few TUs can be classified as WCUs, we hypothesise that global rankings undervalue the merit of the types of institutions that utilise either fewer customised indicators or none at all. In particular, our starting hypothesis is that TUs are generally penalised in the global rankings due to the greater orientation of these league tables towards basic research, to the detriment of university-industry collaboration and a stronger entrepreneurial culture. This is contradictory when at the same time universities are increasingly expected to engage in more interactions with industrial and regional partners (Jongbloed, Enders and Salerno, 2008), and those engaged in applied research activities are compelled to produce economically useful knowledge with industrial relevance (Berbegal-Mirabent et al., 2015). For this reason, university leaders, governments and stakeholders – students, labour market actors, companies and industries – cannot use these global rankings to classify HEIs in general, nor TUs in particular, in terms of entrepreneurial universities or interaction with industry in the knowledge economy (Shattock, 2009).

In order to overcome some of the methodological limitations of global rankings in this respect, ranking editors started to produce tailored university lists by field or discipline. For example, ARWU annually produces rankings by fields of knowledge (Natural Sciences and Mathematics, Engineering, Technology and Computer Science, Life and Agricultural Sciences, Clinical Medicine and Pharmacy, and Social Science), and by subject (Mathematics, Physics, Chemistry, Computer Science, Economic/Business, Chemical Engineering, Civil Engineering, Electrical & Electronic Engineering, Energy Science & Engineering, Environmental Science & Engineering, Materials Science and Engineering, and Mechanical Engineering). THE Ranking also publishes global rankings by subject. In this case, the available lists are the following: Arts & Humanities, Business & Economics, Clinical, preclinical & health, Computer Science, Life Sciences, Physical Sciences, and Social Sciences, and Engineering & Technology. These specialised rankings operate by adapting the overall ranking weights to the research particularities of each field. However, these league tables are not intended to rank TUs, but determine whether to classify universities with a research activity in each of the subjects. That is to say, the World University Rankings by subject (engineering and technology) do not essentially constitute a TUs ranking.

In this study we precisely identify and analyse the TUs included in the THE Ranking (2017) to determine whether the hypothesis of horizontal differentiation (comprehensive versus technical universities) is supported by empirical results. The aim of this article is twofold: firstly, to demonstrate that TUs show a distinct behaviour (at least in the THE ranking) and that within this type of institution we can find different profiles that can be explained in terms of different strategic groups or clusters of TUs. Secondly, we aim to show how their performance might change if some slight changes are applied to the main scores' weights used by the THE ranking editors to give more relevance to industry income criteria through the simulation of several scenarios. The paper ends with a discussion of policy implications and some concluding remarks.

## 2. Data and Method

### 2.1. Data gathering

The first stage consisted of identifying TUs. We selected as a starting pool the available data in the 2017 edition of the Times Higher Education World University Rankings (THE-WUR) (https://www.timeshighereducation.com/world-university-rankings/2017/-

world-ranking). The reason for selecting this particular university ranking was that it included one dimension related to industry income. It is considered a key performance dimension for TUs not covered by other comprehensive global rankings, such as the QS World University Rankings (https://www.topuniversities.com), the Academic Ranking of World Universities (ARWU) (http://www.shanghairanking.com), the CWTS Leiden Ranking (http://www.leidenranking.com) and the Ranking Web of World Universities (http://www.webometrics.info).

Among the total of 981 worldwide universities listed online by the THE ranking in 2017, we identified those universities with a strong technical component. There is no unique definition of a technical university and it is currently a controversial concept under validity. A TU can be defined as an institution of advanced engineering and scientific research that specialises in science, engineering and technology or various types of technical fields. However, the definition or concept of TU is not homogeneous across countries, because it is a cumbersome task to delimit which HEIs should be considered TUs and which should not (De la Torre and Perez-Esparrells, 2017).

In this study we consider as TUs all those universities that contain the words "technical", "technology" or "polytechnic" in their official institutional names (excluding institutions focused on vocational training and those universities that have a faculty of engineering but to not specialise in the aforementioned fields). Additionally, we included all those universities for which, despite not having the aforementioned words in their titles, an important part of their syllabi was technical subjects. To do this, the first author manually checked the online available syllabi on the corresponding official university websites, labelling each institution as a technical university if 40-50% of the subject mix was grounded in technical matters and engineering disciplines (for example, University of Twente). After this, the initial set of institutions obtained was revised by the second author. The inter-rater disagreement was low (<3% of institutions) and related to specific institutions. A more problematic case was Stanford University, an institution that is outstanding in Computer Science but its overall subject mix was under the threshold established. In the end, the institution was excluded. Finally, a second iteration was performed and a total agreement concordance was achieved.

This process resulted in the identification of 137 TUs (14% of all the universities provided by the THE ranking in 2017). For each institution, all the scores on each of the five main dimensions (teaching, research, citations, international outlook and, especially, industry income) were gathered to build a simple tech ranking called TU R (see Appendix I). A brief explanation of the methodology and scope of each dimension, as computed by the THE Ranking by subject: engineering and technology, is offered in Appendix II.

## 2.2. Statistical analysis

Firstly, the descriptive statistics (average, median) of each score dimension was obtained to characterise the sample. Secondly, in order to reveal the relation of industry income to the remaining dimension scores, a correlation test was performed on all five dimensions as well. Each dimension score consisted of a dimensioned value of between 0 and 100, and Pearson ($\alpha < 0.1$) was applied. Thirdly, a cluster analysis (k-means) was applied to the whole sample with the purpose of typifying potentially different technical university profiles. Five predefined classes were selected (to match them to each of the THE's ranking score dimensions: Research, Citations, Teaching, Industry Income and International Outlook), and the clustering criterion was Determinant (W), with a random initial partition and 10 repetitions. Both data gathering and data analysis were performed during January 2017.

### 2.3. Ranking weights simulation

Finally, the global score weights were simulated to obtain alternative scenarios: a "soft" scenario (1), where Research and Citations were altered from 30% to 27.5%, and Industry Income from 2.5% to 7.5%; a "strong" scenario (2), in which Research and Citations were both altered from 30% to 25% and Industry Income from 2.5% to 12.5%; and, finally, a "tech" scenario (3), in which the citations were decreased to 27.5%, whereas Industry was increased to 5% (this scenario corresponds with that used by the World University Rankings 2016-2017 by subject: engineering and technology methodology). Both the original and the simulated weight scores are shown in Table 1.

**Table 1. Summary of the original and simulated weight scores**

The purpose of these scenarios was to readjust the universities' positions by giving a more specific weight to the Industry Income dimension by decreasing the relative weights that corresponded with Research and Citations. In the "strong" scenario (scenario 2) Research and Citations were reduced equally by 5%, transferring this percentage to Industry Income. Scenario 1 was softer, with only Research and Citations reduced by 2.5% each, and the Industry Income score was increased by a corresponding 5%. The third scenario enabled us to check the accuracy of the weights already used by the World University Rankings 2016-2017 by subject (engineering and technology), which comprised General Engineering, Electrical and Electronic Engineering, Mechanical and Aerospace Engineering, Civil Engineering, and Chemical Engineering for a total of 101 worldwide universities.

Note that in scenario 3 the weights assigned to Research were not altered and the rankers moved the weights only from Citations to Industry Income. In our proposed scenarios 1 and 2, the exercise was precisely to carry out a "trade-off" between the traditional research indicators – publications and citations – and the knowledge transfer indicator, namely Income Industry, which is a more entrepreneurial indicator due to the growing importance of university-business collaboration. (For an extensive discussion about the relevance of the university-industry linkages, see Plewa et al., 2013.)

## 3. Results

Once all the institutions with a strong technological component from the THE had been identified, we generated a global ranking for TUs, which is available in Appendix I (sorted by the TU Ranking parameter, TU R). After this, a statistical analysis to describe TUs' performances and the simulation of several ranking scenarios according to Industry Income weight scores were applied, which are fully described in the following sections.

### 3.1 Performance of Technical Universities according to THE ranking Scores

The score in which TUs perform higher on average corresponds with Industrial Income ($\bar{x}$ = 57.4), followed by citations ($\bar{x}$ = 44.9). Conversely, Research ($\bar{x}$ = 26.8) and Teaching ($\bar{x}$ = 30.5) clearly stand out as the weaker dimensions. We can check the average differences across score dimensions through a box-and-whisker plot (Figure 1).

**Figure 1. Statistical performance of dimension scores for Technical Universities**

Nonetheless, we must emphasise the existence of statistical outliers in the sample. In order to detect them, a Dixon test for outliers was applied (significance level: 5%; p-value: 0.128; Number of Monte Carlo simulations: 1,000,000). In Table 2 we can observe the number of outliers according to each of the five dimensions, labelling each of the universities with the ranking position achieved in Appendix I (TU Ranking).

**Table 2. Statistical outliers broken down by THE Dimensions**

The Research score contained the highest number of outliers (9) followed by International Outlook and Teaching (5 each), and Citations (4). Finally, the Industry Income variable presented no outliers. Additionally, we could check how the outliers were grouped in a few institutions. In fact, only 12 out of the 137 universities had outliers in at least one dimension. The universities ranked first (California Institute of Technology) and second (Massachusetts Institute of Technology) exhibited outlier behaviour in their Teaching, Research and Citations scores, whereas the university ranked third (Imperial College London) behaved as an outlier in Teaching, International Outlook, Research and Citations.

The statistical performance of the dimensions can be clearly pictured if we compare technical universities with non-technical universities (Table 3).

**Table 3. Statistical comparison (mean value) of Technical Universities and Non-Technical Universities**

As we can observe, the Teaching and Research scores exhibited similar mean values for TUs and Non-TUs. However, the International Outlook, Citations and especially Industry Income showed important differences. In order to check whether these differences were statistically significant, a k-sample comparison of variances (Levene's test [Mean]) was performed (Table 4), confirming the statistical differences (p-value < 0,0001; α = 0.05) in the Industry Income score according to the university's class (TU or Non-TU).

**Table 4. Levene's test (mean) between Technical Universities and Non-Technical Universities**

The prominence of Industry Income scores for non-technical universities is shown in Figure 2. As we can observe, 50% of the non-technical universities achieved lower scores (below 50 score points), although we found a group of high-performers (29 universities out of the 841 non-technical universities of the sample surpassed 90 score points). Despite this, Figure 2 indicates the overall difference on Industry Income performance in the case of TUs.

**Figure 2. Industrial income for Technical and non-Technical Universities**

Otherwise, a significant difference between the Research (25.7) and Citation (50.1) average scores in the case of TUs was found (see Table 2), although these two dimensions were strongly related each other. The reason for this is the method used by THE ranking to generate scores from raw data. First, the institution that obtains the highest score in each dimension is labelled with a score equal to 100 and, secondly, the remaining institutions are scored correspondingly.

Therefore, the difference in percentage terms of universities with respect to the top performer are likely to be higher in Citations (impact) than Research (productivity) due to the more skewed distribution of citation data. This is especially true if we consider the top technical institutions (Caltech and MIT), which correspond with WCU in terms of an outstanding research output.

However, the Research and Citation scores correlated strongly (r = 0.74), that is, those technical institutions that published more contributions were the same ones that received a higher number of citations (Table 5; up). In this case, we can observe a weaker correlation of Industry Income, both with Research (0.57) and especially with Citations (r= 0.35). Curiously enough, the Research and Teaching scores achieved the strongest correlation (r= 0.94), presumably due to the influence of the expert surveys used to obtain these indicators (reputation is biased towards research excellence), whereas the Industry Income and International Outlook scores had the weakest correlation (r= 0.13). If we compare these results with a correlation matrix of all 978 universities (Table 5; bottom), we can see that the correlation of Industry Income with both Research and Citations decreases (0.43 and 0.21 respectively).

**Table 5. Pearson correlation matrix for THE ranking scores: Technical universities (up) and all universities (bottom)**

Beyond the general correlations calculated, we found strategic groups of universities exhibiting specific behaviours, highlighting in some scores and backgrounding in others. In order to detect these potential TUs' profiles, a k-means clustering algorithm was applied. The composition of each of the five clusters identified, as well as the country of each institution, is available in Appendix III. A radial representation of each cluster is displayed in Figure 3.

**Figure 3. Technical Universities clusters according to THE ranking scores**

The first cluster comprises 20 universities (all of them top ranked in the global score). The distribution of the remaining clusters is quite balanced (the second cluster consists of 29 institutions; the third cluster of 26; the fourth of 28; and fifth of 34 universities). The average rank position of TUs assigned to each of the clusters and a proposal to describe each cluster is available in Table 3.

**Table 6. Composition of technical universities clusters**

As we can observe, the overall position in the global ranking is determined by affiliation to one or another cluster. Therefore, the top universities (Cluster 1) exhibit a differentiated score-performing pattern, characterised to highlight all score dimensions, except for teaching in some cases. They can classify as WCUs and in that way be defined as global TUs. Cluster 4 shows similar behaviour (comparable shape of Cluster 1), although with generally lower scores. Otherwise, clusters 3 and 5 are also quite similar in shape among themselves, exposing an excellent performance in Industry Outcomes, whereas Cluster 2 is the only one in which the International Outlook score is highlighted. Yet, these results must not be interpreted as taxonomy or even as a typology of TUs, but as a first step of profile characterisation.

The different behaviour of universities within each of the clusters can be additionally observed through the existence or absence of specific correlations between dimensions (Table 7). For example, the correlation between the Teaching and Research scores is high for Cluster 1 ($R_p$= .85), although lower in cluster 5 ($R_p$= .64). Citations and Research are highly correlated in Cluster 1 ($R_p$= .74) but uncorrelated in clusters 4 ($R_p$= .17) and 5 ($R_p$= -.02). With regard to Industry Income, this score exhibits an absence of correlation with all the other scores in all clusters except Cluster 5, where it correlates with Research ($R_p$= .59)

**Table 7. Correlation between dimensions broken down by Cluster**

The results indicated in Cluster 1 are aligned with those previously obtained by De la Torre and Perez-Esparrells (2017) on European TUs, who claim that university groups more aligned with the global rankings criteria are those TUs with more resources (in absolute and relative terms), with a stronger emphasis on PhD programmes in relation to the total teaching activity of the institution, with a higher internationalisation of their human resources (both students and academic staff), and with a more comprehensive subject mix orientation (specially the existence of Health studies and Medicine faculties).

Otherwise, as we can see in Appendix I, the differences between TUs within one cluster and country are greater than among TUs belonging to different countries. The Top-5 countries in the TUs are China (12 TUs), followed by India (11), Japan (11), Germany (10) and the United States (10). The alluvial diagram (Figure 4) indicates the distribution of countries and clusters more comprehensively. Here we can observe how Germany stands out in TUs integrated in Cluster 1 (5 out of 10), China in Cluster 3 (7 out of 12), India in Cluster 4 (5 out of 11) and Japan in Cluster 5 (9 out of 11). Cluster 2, despite being highly distributed, highlights the hosting of all three Australian TUs covered by the THE ranking.

**Figure 4. Distribution of countries broken down by cluster**

In light of the above, it seems senseless to compare institutions among university systems or subsystems within a geographical area (Europe, Asia, and Latin America) if specific profiles, such as in the case of TUs, are not considered.

### 3.2 Scenarios simulated with Industry Income scores

In Table 8, we can see the top 20 TUs according to the overall score provided by the THE ranking (2017 edition). In addition to this, we introduce the simulated score (and the corresponding rank position) that these same universities would have obtained if Research, Citation and Industry Income score had been weighted according to the three different scenarios previously described (soft, strong and tech).

**Table 8. Simulation scenarios of Technical universities rankings**

Interestingly enough, the first top six TUs in the original ranking (Caltech, MIT, Imperial College London, ETH Zurich, École Polytechnique Fédérale de Lausanne and Georgia Institute of Technology) show a decrease in their overall scores when altering score weights. These institutions (all of them belong to the Cluster 1 previously mentioned) exhibit a world-class behaviour (with a strong performance in Research) that distinguishes them from the remaining TUs, as previously mentioned.

In general terms, these top universities essentially maintain their ranking positions (Caltech would have climbed to the first position in the three scenarios proposed). However, as we descend in the ranking list we observe an increasing effect on institutions, despite the existence of some outliers (such as the University of Warwick).

This tail effect is well demonstrated in Figure 5, in which a histogram shows the distribution of the score difference between the original and strong scenario (original score minus strong score). We can observe that the scores of 21 out of 137 technical universities (15.3%) would worsen with the strong scenario. Of these, five universities (all of them belonging to Cluster 1) would lose more than two points (Imperial College, ETH Zurich, École Normale Supérieure, University of Warwick). Conversely, the remaining 116 universities would improve their scores, 17 of them by more than five points.

**Figure 5. Histogram of difference between original and strong scenario scores for technical universities**

We can contrast the rank improvement achieved by technical universities by counting the number of institutions positioned within specific margin positions (Table 9). As we can see, when we alter the score weights, the number of universities slightly increases in each of the ranking sections (especially for the top 400 and top 500 positions in the strong scenario).

**Table 9. Ranking position of technical universities for each simulated scenario**

## 4. Discussion

On the one hand, in recent years rankings have produced homogeneity in university performance but heterogeneity in the mission mix of the institutions, and their strategies still prevail. This growing concern entails the need to separate the typologies of universities (such as a specific ranking for TUs, which is the purpose of this study) and/or to promote different types of university rankings that are capable of measuring other functions, such as third mission rankings (Montesinos et al, 2008). On the other hand, although universities have tried to work out how to protect academic autonomy from all kinds of interference, it is noteworthy that some institutions willingly allow their decisions to become vulnerable to an agenda set by others (Altbach and Hazelkorn, 2017).

The source selection, the TUs identification procedure and the simulated scenarios constitute the main issues to discuss about the validity of the results previously described.

Firstly, among the several global rankings currently available, the authors of this study opted for the THE. This ranking includes the Industry Income score, which supposes an essential dimension to characterise TUs amongst the institutions. Obviously, this fact limits the overall sample of universities to analyse (981 institutions). We would have found additional TUs if we had accessed comprehensive global rankings (for example, the Ranking Web of World Universities covers more than 24,000 HEIs). However, neither the identification of TUs nor the collection of distinctive indicators for TUs was feasible.

Secondly, the procedure to categorise whether an institution is a technical university introduces some inherent errors (due to either false inclusion or exclusion). However, we can say that the most prominent TUs worldwide are among the 137 universities included.  Moreover, the inclusion or exclusion of certain particular institutions does not statistically affect to the main findings of this study.

Thirdly, the scenarios simulated obviously constitute just three of the possible score weight combinations we can figure out. We assume first-mission (teaching) scores are of equal relevance to all universities (all of them have this essential mission). Moreover, due to the globalisation process, most universities have internationalised (both students and teachers) due to internationalisation being an intrinsic feature of the current knowledge economy. Consequently, we decided to keep the importance given by THE ranking to the International Outlook score. However, second-mission scores (represented by Research and Citations) are only one essential aspect of research-intensive institutions (which cover some key disciplines and knowledge fields, such as Medicine, Physics and Chemistry). Nonetheless, TUs do not necessarily place this important weight on research. This is the reason for which we propose a trade-off between Research and Industry Income scores. The three scenarios proposed are realistic (the first softer, the second stronger and the third utilised precisely by THE-ENG) and sufficient enough to properly meet the objective of proving a masking of the merits (in this case measured by Industry Income). The low correlation between Industry Income and Research in TUs demonstrates that they constitute differing dimensions of this type of institution.

In short, universities vary enormously in the extent to which they promote and succeed in commercialising academic research (Geuna and Muscio, 2009). In all events, ranking results should be contextualised since the final positions are prone to be influenced by the anchoring effect (Bowman and Bastedo, 2011), underlying factors (Safón, 2013) and statistical noise or exceptional events (Piro and Sivertsen, 2016).

## 5. Conclusions

In this study a first-world technical university ranking composed of 137 universities belonging to the THE ranking was presented. Among these institutions, we can discover the national technological flagships in their home countries that bring new advances for scientific and technology transfers and the majority of commercialisation of university research results.

The results demonstrate a distinct statistical behaviour of TUs, characterised by moderate-to-high scores in Industry Income and a low performance in Research scores if compared with the remaining non-Technical universities of the Top-800 sample in THE ranking.

Given the weights applied to each of the dimensions (30% Research; 30% Citations; 30%; Teaching; 7.5% International Outlook; and 2.5% Industry Income), the original ranking manifestly diminishes the potential of TUs' performances. When simulating alternative scenarios (especially the "soft" and "strong" scenarios), in which the Industry Income weight is slightly increased (because of decreasing Research and Citations), the results show an overall increase in scores of the majority of TUs. A total of 83.2% of TUs (114) increase their overall scores, some of them significantly (17 universities would benefit from an increase of more than five points in their final overall score within the strong scenario).

However, this masking effect does not affect all TUs equally, since we find up to five clusters with a differentiated performance pattern. The first cluster comprises WCUs that over-perform in all dimension scores. Their behaviour is similar to other research-intensive world-class universities, largely due to the inclusion of both Medicine and Physics in some institutions. The scenarios proposed prejudice the scores of these top institutions. The second cluster comprises TUs with a special emphasis on international outlook. The effect of the simulated scenarios for these universities was moderate. The third and fifth clusters capture TUs with national and local impacts respectively. The effect of the simulated scenarios on these clusters is noteworthy. Finally, the fourth cluster consists of TUs that achieve a huge research impact. The effect of simulated scenarios on these institutions was less pronounced than that obtained for clusters 3 and 5, which were more industry oriented. In any case, the consequences of implementing a ranking with different weights towards industry income have revealed a massive impact on TUs, especially on those that have not yet attained world-class status.

In view of the results obtained (which demonstrate a particular behaviour for this type of institution), a different yardstick should be used to measure a higher education policy perspective. In line with this, the authors of this study indeed aim to finish with some recommendations and policy implications at several levels (providers, institutions and governments).

*Ranking providers' level*. Over and above their current offering, they need to generate new rankings that are especially tailored for TUs or adapt the old ones by including new criteria on knowledge transfer and innovation to measure the different aspects offered by institutions with more technical and entrepreneurial profiles. These innovative rankings with new criteria may not only serve to measure the specificity of TUs, but may also be sensitive to the different profiles of TUs already identified in this study.

*Institutional level*. The existence of different ways of performing rankings would strategically serve to better define new strategies in order to ensure that institutions with a traditional mission that is oriented towards the industry achieve excellence (Hazelkorn et al., 2014). Additionally, they can help to establish better definition for funding schemes: to nurture their scientific and technological leadership; to guarantee future prosperity and innovative strength in their respective countries; to boost technological research to foster solutions in major world issues; and to acquire adequate critical mass in key sectors, such as energy, green technologies, nano-engineering, the material sciences, biomedicine, space and transport.

*Government level*. Public policies must include funding and the establishment of incentive schemes for those TUs driven by global rankings and which are striving to keep their world-class status. However, at the same time, governments must not forget those other TUs that choose to remain strictly national, regional or local, and that have more defined regional boundaries and missions. There are no unique and easy solutions to such contradictions emerging from ranking regimes, especially for technical universities. What is clear is that TUs are in a process of rapid change and transformation in the landscape of higher education. In addition, policymakers must resolve these predicaments through incentive systems that comprehend both international and domestic needs.

To sum up, although universities are expected to excel at both basic and applied research activities, in light of the evidence it is apparent that the current global rankings are skewed to productivity research (even to reputation) and the existing methodologies do not support the new emphasis on research outputs with clear practical uses. With the previous analysis we have evidenced the masking of the merits of TUs in one of the global rankings, the THE ranking. However, industry income undoubtedly constitutes only one amongst the many possible proxies for TUs.

Further studies should be able to facilitate the gathering of complementary indicators (for instance, patents, technology licensing, consulting services and advisory projects, launching technology-oriented start-ups and spinoffs) to paint a clearer portrait of TUs, regardless the already existing "rules of the game" in global rankings. The use of qualitative comparative analysis (QCA) (Ragin, 1997; Roig-Tierno, Gonzalez-Cruz and Llopis-Martinez, 2017) as a complement to traditional correlation methods is also advisable as a method to be used in future research.

Finally, it would be worth investigating how universities clustered in each group define themselves. This would make it possible to scrutinise for similarities, not only in terms of performance but also in the mission-mix definition.

## References

Altbach, P. G., 2007. Globalization and the university: Realities in an unequal world. In International handbook of higher education (pp. 121–139). Springer Netherlands.

Altbach, P. G., Salmi, J. (Eds.), 2011. The road to academic excellence: The making of world-class research universities. World Bank Publications.

Altbach, P.G., Hazelkorn, E. 2017. Pursuing Rankings in the Age of Massification: For Most-Forget About it. International Higher Education, 89 (8-10), https://doi.org/10.6017/ihe.2017.89.9834

Berbegal-Mirabent, J., Sánchez García, J. L., Riberio-Soriano, D.E., 2015. University– industry partnerships for the provision of R&D services. Journal of Business Research 68(7), 1407–1413, http://dx.doi.org/10.1016/j.jbusres.2015.01.023.

Berbegal-Mirabent, J., Solé, F., 2012. What are we measuring when evaluating universities' efficiency?, Regional and Sectoral Economic Studies 12(3), 31–46.

Bookstein, F. L., Seidler, H., Fieder, M., Winckler, G., 2010. Too much noise in the Times Higher Education rankings. Scientometrics, 85(1), 295–299, https://doi.org/10.1007/s11192-010-0189-5

Bowman, N. A., Bastedo, M. N., 2011. Anchoring effects in world university rankings: Exploring biases in reputation scores. Higher Education, 61(4), 431–444, https://doi.org/10.1007/s10734-010-9339-1

Butler, D., 2010. University rankings smarten up. Nature 464, 16-17, https://doi.org/10.1038/464016a
Caldera, A., Debande, O., 2010. Performance of Spanish universities in technology transfer: An empirical analysis. Research Policy 39(9), 1160–1173, http://dx.doi.org/10.1016/j.respol.2010.05.016
Cheng, Y., Wang, Q., Liu, N. C., 2014. How world-class universities affect global higher education. In How World-Class Universities Affect Global Higher Education (pp. 1–10). Sense Publishers.
De la Torre, E.M., & Perez-Esparrells, C. (2017). New Strategies of European Technical Universities in the Emerging Competitive Environment of Global Rankings. In S. Dent, L. Lane and T. Strike Collaboration, Communities and Competition (pp. 51-71). Rotterdam: SensePublishers, https://doi.org/10.1007/978-94-6351-122-3_4
Douglass, J.A., 2016. The New Flagship University. Changing the Paradigm from Global Ranking to National Relevancy. New York: Palgrave Macmillan.
Frenken, K., Heimeriks, G. J., & Hoekman, J., 2017. What drives university research performance? An analysis using the CWTS Leiden Ranking data. Journal of Informetrics, 11(3), 859-872, https://doi.org/10.1016/j.joi.2017.06.006
Gomez-Sancho, J.M., Perez-Esparrells, C., 2012. International Higher Education rankings at a glance: how to valorize the research in Social Sciences and Cultural Studies. In López-Varela, A. (Ed.). Social Sciences and Humanities. Applications and Theories (pp 355–374). InTech, London.
Harvey, L., 2008. Rankings of higher education institutions: a critical review. Quality in Higher Education, 14(3), 187–207, https://doi.org/10.1080/13538320802507711
Hazelkorn, E., 2015. Rankings and the Reshaping of Higher Education. The Battle for World-Class Excellence. Basingstoke: Palgrave MacMillan.
Hazelkorn, E., Loukkola, T., Zhang, T., 2014. Rankings in Institutional Strategies and Processes: impact or illusion?. Brussels: European Universities Association.
Hsu, D. W., Shen, Y. C., Yuan, B. J., Chou, C. J., 2015. Toward successful commercialization of university technology: Performance drivers of university technology transfer in Taiwan, Technological Forecasting and Social Change 92, 25–39, http://dx.doi.org/10.1016/j.techfore.2014.11.002.
Johnes, J., 2018. University rankings: What do they really show?. Scientometrics, pp. 1-22, https://doi.org/10.1007/s11192-018-2666-1
Jongbloed, B., Enders, J., & Salerno, C., 2008. Higher education and its communities: Interconnections, interdependencies and a research agenda. Higher education, 56(3), 303–324, https://link.springer.com/article/10.1007/s10734-008-9128-2
Laredo, P., 2007. Revisiting the third mission of universities: Toward a renewed categorization of university activities?. Higher education policy, 20(4), 441–456, https://doi.org/10.1057/palgrave.hep.8300169
Margison, S., 2017. Do Rankings Drive Better Performance?. International Higher Education, 89 (6-8), https://doi.org/10.6017/ihe.2017.89.9833
Minguillo, D., Tijssen, R., Thelwall, M., 2015. Do science parks promote research and technology? A scientometric analysis of the UK. Scientometrics, 102(1), http://dx.doi.org/10.1007/s11192-014-1435-z
Moed, H.F., 2017. A critical comparative analysis of five world university rankings. Scientometrics, 110(2), 967-990, https://doi.org/10.1007/s11192-016-2212-y
Montesinos, P., Carot, J. M., Martinez, J. M., & Mora, F., 2008. Third mission ranking for world class universities: Beyond teaching and research. Higher Education in Europe, 33(2–3), 259–271, https://doi.org/10.1080/03797720802254072
Olcay, G. A., & Bulu, M., 2017. Is measuring the knowledge creation of universities possible?: A review of university rankings. Technological Forecasting and Social Change, 123, 153-160, https://doi.org/10.1016/j.techfore.2016.03.029
Orduna-Malea, E., 2012. Propuesta de un modelo de análisis redinformétrico multinivel para el estudio sistémico de las universidades españolas (2010). Valencia: Universitat Politècnica de València.

Perez-Esparrells, C., De la Torre, E. M., 2017. New strategies of European technical universities in the emerging competitive environment of global rankings. In Strike, T. (Ed.) Collaboration, communities and competition: international perspectives from the academy, Rotterdam: Sense publishers–Springer.

Piro, F. N., Sivertsen, G., 2016. How can differences in international university rankings be explained?. Scientometrics, 109(3), 2263–2278, https://doi.org/10.1007/s11192-016-2056-5

Plewa, C., Korff, N., Johnson, C., Macpherson, G., Baaken, T., Rampersad, G. C., 2013. The evolution of university-industry linkages. A framework, Journal of Engineering and Technology Management 30, 21–44, http://dx.doi.org/10.1016/j.jengtecman.2012.11.005

Safón, V., 2013. What do global university rankings really measure? The search for the X factor and the X entity. Scientometrics, 97, 223–244, https://doi.org/10.1007/s11192-013-0986-8

Sarrico, C. S., Rosa, M. J., Teixeira, P. N., Cardoso, M. F., 2010. Assessing quality and evaluating performance in higher education: Worlds apart or complementary views?, Minerva 48(1), 35–54, http://dx.doi.org/10.1007/s11024-010-9142-2.

Shattock, M., 2009. Entrepreneurialism in universities and the knowledge economy. London, UK: Society for Research into Higher Education and Open University Press.

Ragin, C., 1987. The comparative method. Berkeley: University of California Press.

Rauhvargers, A., 2011). Global rankings and their impact. EUA report on rankings 2011. Brussels: The European University Association.

Ribeiro-Soriano, D.E. & Berbegal-Mirabent, J., 2017. Disseminating scientific research: a double-edged sword?. Knowledge Management Research & Practice, 15(3), 380-390, https://doi.org/10.1057/s41275-017-0070-x

Roig-Tierno, N., Gonzalez-Cruz, T.F., Llopis-Martinez, J., 2017. An overview of qualitative comparative analysis: A bibliometric analysis. Journal of Innovation and Knowledge 2(1), 15-23, https://doi.org/10.1016/j.jik.2016.12.002.

Van Raan, A. F., 2005. Fatal attraction: Conceptual and methodological problems in the ranking of universities by bibliometric methods. Scientometrics, 62(1), 133–143, http://dx.doi.org/10.1007/s11192-005-0008-6

**Appendix I. Technical Universities master list (THE Ranking, 2017 edition)**

| TU R | Global R | Cluster | Technical University (TU) | Overall | Citations | Industry income | International Outlook | Research | Teaching | Country |
|---|---|---|---|---|---|---|---|---|---|---|
| **1** | **2** | 1 | California Institute of Technology | **94.33** | 99.8 | 90.8 | 63.4 | 95.7 | 95.5 | USA |
| **2** | **5** | 1 | Massachusetts Institute of Technology | **93.38** | 99.9 | 88.4 | 85.6 | 92.3 | 90.3 | USA |
| **3** | **8** | 1 | Imperial College London | **90.02** | 97.3 | 67.5 | 96.5 | 86.6 | 86.4 | UK |
| **4** | **9** | 1 | ETH Zurich – Swiss Federal Institute of Technology Zurich | **89.26** | 92.5 | 63.7 | 98.1 | 93.7 | 81.5 | Switzerland |
| **5** | **30** | 1 | École Polytechnique Fédérale de Lausanne | **76.79** | 96.5 | 69.8 | 98.6 | 66.1 | 62.9 | Switzerland |
| **6** | **33** | 1 | Georgia Institute of Technology | **76.26** | 90.8 | 62.3 | 72.8 | 79.2 | 60.8 | USA |
| **7** | **40** | 1 | KU Leuven | **73.79** | 90.1 | 99.8 | 67.4 | 73.7 | 57 | Belgium |
| **8** | **46** | 1 | Technical University of Munich | **71.55** | 82 | 100 | 66.6 | 70.5 | 61 | Germany |
| **9** | **49** | 1 | Hong Kong University of Science and Technology | **71.09** | 91.2 | 62 | 82.8 | 66.7 | 53.2 | China |
| **10** | **54** | 1 | Nanyang Technological University | **69.97** | 90.7 | 93.5 | 95.7 | 60.2 | 50.6 | Singapore |
| **11** | **59** | 1 | Delft University of Technology | **67.86** | 67.3 | 99.9 | 85.9 | 72.9 | 56.2 | Netherlands |
| **12** | **66** | 1 | École Normale Supérieure | **65.79** | 85.8 | 40 | 72.2 | 52.3 | 59.8 | France |
| **13** | **78** | 1 | RWTH Aachen University | **62.98** | 71.3 | 99.4 | 53.4 | 63.7 | 53.3 | Germany |
| **14** | **82** | 1 | Technical University of Berlin | **62.03** | 74.3 | 98 | 60.8 | 59.3 | 49.8 | Germany |
| **15** | **82** | 2 | University of Warwick | **61.97** | 80.9 | 40.8 | 91.4 | 52.6 | 46.8 | UK |
| **16** | **89** | 1 | Korea Advanced Institute of Science and Technology (KAIST) | **61.29** | 78.5 | 100 | 34.3 | 53.2 | 55.7 | South Korea |
| **17** | **104** | 1 | Pohang University of Science and Technology | **59.57** | 79.2 | 99.6 | 34.2 | 48.7 | 53.8 | South Korea |
| **18** | **116** | 1 | École Polytechnique | **58.64** | 67.2 | 71.9 | 92.3 | 40.6 | 58.6 | France |
| **19** | **144** | 1 | Karlsruhe Institute of Technology | **55.79** | 76.4 | 98.3 | 59.5 | 45.2 | 41.3 | Germany |
| **20** | **153** | 3 | University of Science and Technology of China | **54.73** | 74.7 | 77.6 | 24.2 | 43.4 | 51.8 | China |
| **21** | **153** | 1 | University of Twente | **54.68** | 70.8 | 84.6 | 83.6 | 46.7 | 36.8 | Netherlands |
| **22** | **159** | 2 | KTH Royal Institute of Technology | **54.09** | 67.4 | 51.1 | 83.3 | 45.3 | 42.5 | Sweden |
| **23** | **164** | 1 | TU Dresden | **53.53** | 70.2 | 95.7 | 49.4 | 45.6 | 42.3 | Germany |
| **24** | **176** | 2 | Technical University of Denmark | **52.50** | 80.2 | 60.6 | 85.8 | 30 | 38.3 | Denmark |
| **25** | **177** | 2 | Eindhoven University of Technology | **52.36** | 71.7 | 56.9 | 72 | 43.5 | 36.6 | Netherlands |
| **26** | **190** | 2 | Newcastle University | **51.35** | 81 | 38.7 | 84.9 | 32.1 | 33.6 | UK |
| **27** | **192** | 2 | Hong Kong Polytechnic University | **51.19** | 69.9 | 47.9 | 79.7 | 41.8 | 35 | Germany |
| **28** | **201-250** | 3 | University of Stuttgart | **48.10** | 53.4 | 100 | 47.2 | 44.5 | 42.3 | Germany |
| **29** | **201-250** | 2 | Aalto University | **48.05** | 75 | 52.5 | 69.6 | 29 | 34.4 | Finland |
| **30** | **201-250** | 2 | École Normale Supérieure de Lyon | **47.82** | 64.5 | 34.9 | 68.4 | 30.4 | 44.5 | France |
| **31** | **201-250** | 2 | Queensland University of Technology | **47.43** | 67.4 | 58.9 | 77.9 | 37.9 | 28.4 | Australia |
| **32** | **201-250** | 2 | Polytechnic University of Milan | **47.22** | 76.4 | 59.3 | 52.6 | 30.3 | 32.6 | Italy |
| **33** | **201-250** | 3 | Technical University of Darmstadt | **46.59** | 52.5 | 95.1 | 54.7 | 43 | 38.2 | Germany |
| **34** | **251-300** | 3 | Tokyo Institute of Technology | **46.25** | 41.7 | 64.1 | 32.9 | 50.5 | 48.4 | Japan |
| **35** | **251-300** | 2 | Chalmers University of Technology | **46.22** | 61.9 | 70.8 | 75.8 | 28.5 | 38.8 | Sweden |
| **36** | **251-300** | 3 | Rensselaer Polytechnic Institute | **45.53** | 67.3 | 79.1 | 43.1 | 33.5 | 33.6 | USA |
| **37** | **251-300** | 2 | University of Technology Sydney | **45.05** | 73.8 | 44.2 | 92.3 | 22.5 | 27.1 | Australia |
| **38** | **251-300** | 2 | Norwegian University of Science and Technology | **44.58** | 65.5 | 46.5 | 65.3 | 32.8 | 30.1 | Norway |
| **39** | **251-300** | 4 | Virginia Polytechnic Institute and State University | **44.53** | 62.1 | 44.1 | 31.4 | 39.9 | 34.9 | USA |
| **40** | **251-300** | 2 | Vienna University of Technology | **44.10** | 57.7 | 70.9 | 79.1 | 25.5 | 38.1 | Austria |
| **41** | **301-350** | 2 | Technion Israel Institute of Technology | **43.22** | 55.9 | 38.7 | 62.5 | 36.1 | 33.2 | Israel |
| **42** | **301-350** | 4 | Gwangju Institute of Science and Technology | **43.15** | 51.9 | 44.9 | 35.2 | 39.9 | 39.5 | South Korea |
| **43** | **301-350** | 3 | Moscow Institute of Physics and Technology | **42.70** | 40.8 | 98.7 | 50 | 33.1 | 47.7 | Russia |
| **44** | **301-350** | 4 | Technical University of Dortmund | **41.05** | 57.1 | 48.7 | 41.5 | 29.8 | 35.5 | Germany |
| **45** | **351-400** | 3 | Leibniz University of Hanover | **40.20** | 43.2 | 62 | 44.5 | 37.1 | 37.4 | Germany |
| **46** | **351-400** | 4 | Indian Institute of Technology Bombay | **39.54** | 52.8 | 46.9 | 19.2 | 31.1 | 39.2 | India |
| **47** | **351-400** | 2 | Swinburne University of Technology | **38.64** | 60.3 | 35.5 | 74.5 | 24 | 22.9 | Australia |
| **48** | **351-400** | 4 | Toyota Technological Institute | **38.62** | 78.6 | 59 | 24.1 | 17 | 22.2 | Japan |
| **49** | **351-400** | 2 | Polytechnic University of Turin | **38.35** | 66.8 | 49.4 | 48.5 | 19 | 25.8 | Italy |
| **50** | **351-400** | 2 | Graz University of Technology | **38.21** | 60 | 67.8 | 71.2 | 15.4 | 28.5 | Austria |
| **51** | **401-500** | 4 | Marche Polytechnic University | **37.31** | 85.8 | 37.5 | 28.1 | 7.2 | 21.2 | Italy |
| **52** | **401-500** | 3 | National Taiwan University of Science and Technology | **36.91** | 38.5 | 70.1 | 32.3 | 40.5 | 30.1 | Taiwan |
| **53** | **401-500** | 2 | Cyprus University of Technology | **36.03** | 70.3 | 34 | 67 | 13.1 | 17.1 | Cyprus |
| **54** | **401-500** | 3 | Indian Institute of Technology Delhi | **35.51** | 47 | 71.7 | 15.9 | 23.9 | 37.5 | India |
| **55** | **401-500** | 2 | Stevens Institute of Technology | **34.33** | 52.2 | 33.7 | 77.3 | 18.3 | 21.8 | USA |
| **56** | **401-500** | 3 | Indian Institute of Technology Madras | **33.88** | 41.2 | 81.1 | 17.5 | 22.3 | 38.3 | India |
| **57** | **401-500** | 4 | Polytechnic University of Catalonia | **33.63** | 51.2 | 41.5 | 51.4 | 17.5 | 27.1 | Spain |
| **58** | **401-500** | 2 | King Fahd University of Petroleum and Minerals | **33.49** | 36.7 | 92.8 | 86.8 | 16.8 | 28.7 | Saudi Arabia |
| **59** | **401-500** | 2 | Federico Santa María Technical University | **33.33** | 69 | 41 | 54.7 | 10.8 | 14.2 | Chile |
| **60** | **401-500** | 3 | Huazhong University of Science and Technology | **32.98** | 43.6 | 68 | 19.9 | 24.7 | 31 | China |
| **61** | **401-500** | 3 | Indian Institute of Technology Kanpur | **32.58** | 38.4 | 98.8 | 17.9 | 22.9 | 34.6 | India |
| **62** | **501-600** | 2 | Khalifa University of Science, Technology and Research | **32.20** | 43.2 | 45.8 | 93.3 | 12.1 | 24.9 | UAE |
| **63** | **501-600** | 4 | Indian Institute of Technology Roorkee | **32.09** | 55.1 | 64.3 | 15.2 | 15.8 | 26.9 | India |
| **64** | **501-600** | 4 | Tomsk Polytechnic University | **31.73** | 43.2 | 59.1 | 40.6 | 16.2 | 31.3 | Russia |
| **65** | **501-600** | 4 | Polytechnic University of Valencia | **31.49** | 43.9 | 44.3 | 41.9 | 24.8 | 22.1 | Spain |
| **66** | **501-600** | 2 | New Jersey Institute of Technology | **31.42** | 47.9 | 47.6 | 55.1 | 17.8 | 21.3 | USA |
| **67** | **501-600** | 4 | Indian Institute of Technology Kharagpur | **30.91** | 44.5 | 49.5 | 16 | 19.4 | 31 | India |
| **68** | **501-600** | 3 | Monterrey Institute of Technology and Higher Education | **30.84** | 36.5 | 93.5 | 67.6 | 16.6 | 25 | Mexico |
| **69** | **501-600** | 2 | Auckland University of Technology | **30.82** | 45.9 | 33.9 | 94.8 | 12.5 | 17.8 | New Zealand |
| **70** | **501-600** | 3 | Lappeenranta University of Technology | **30.75** | 28.7 | 61.5 | 55.9 | 26.3 | 28.4 | Finland |
| **71** | **501-600** | 3 | Sharif University of Technology | **30.51** | 37.6 | 85.3 | 17.9 | 27.4 | 25.1 | Iran |
| **72** | **501-600** | 2 | Tampere University of Technology | **30.48** | 44.3 | 64.6 | 56.1 | 14.9 | 23 | Finland |
| **73** | **501-600** | 4 | East China University of Science and Technology | **30.32** | 51.9 | 60.7 | 25.6 | 15.2 | 22.5 | China |
| **74** | **501-600** | 2 | Missouri University of Science and Technology | **29.64** | 34 | 45.4 | 65.3 | 24.7 | 20 | USA |
| **75** | **501-600** | 3 | Harbin Institute of Technology | **29.50** | 24.4 | 97.7 | 27.9 | 27.7 | 31.1 | China |
| **76** | **501-600** | 4 | Iran University of Science and Technology | **28.92** | 41.8 | 50.7 | 13.9 | 23.7 | 23.2 | Iran |
| **77** | **501-600** | 4 | Warsaw University of Technology | **28.72** | 59 | 36.7 | 21.5 | 9.9 | 18.4 | Poland |
| **78** | **501-600** | 3 | Istanbul Technical University | **28.66** | 32.4 | 100 | 24.8 | 24.6 | 24 | Turkey |
| **79** | **501-600** | 4 | Indian Institute of Technology Guwahati | **27.57** | 42.1 | 35.4 | 14.6 | 16.6 | 26.6 | India |
| **80** | **601-800** | 3 | Isfahan University of Technology | **27.03** | 38.4 | 89.8 | 18.9 | 20 | 19.5 | Iran |
| **81** | **601-800** | 4 | Middle East Technical University | **27.02** | 31.4 | 58.8 | 27.8 | 19.5 | 27.3 | Turkey |
| **82** | **601-800** | 4 | Czech Technical University in Prague | **26.97** | 36.4 | 38.5 | 42.4 | 20.8 | 18.9 | Czech Republic |
| **83** | **601-800** | 3 | South China University of Technology | **26.85** | 39.5 | 67.9 | 18.6 | 19.4 | 20.3 | China |
| **84** | **601-800** | 2 | Jordan University of Science and Technology | **25.97** | 50.2 | 33.2 | 62.4 | 7.6 | 10.4 | Jordan |
| **85** | **601-800** | 4 | Amirkabir University of Technology | **25.78** | 35.5 | 59.9 | 15.4 | 18.9 | 22.7 | Iran |
| **86** | **601-800** | 4 | Florida Institute of Technology | **25.77** | 28.8 | 36.6 | 60.6 | 13.4 | 25.5 | USA |
| **87** | **601-800** | 4 | Tallinn University of Technology | **25.55** | 44.7 | 44.7 | 42.5 | 11.4 | 14.7 | Estonia |
| **88** | **601-800** | 3 | King Mongkut's University of Technology Thonburi | **25.20** | 42.1 | 80.3 | 28.8 | 11.4 | 16.6 | Thailand |
| **89** | **601-800** | 3 | Dalian University of Technology | **24.97** | 29.3 | 81.5 | 21 | 20.1 | 21.8 | China |
| **90** | **601-800** | 4 | COMSATS Institute of Information Technology | **24.77** | 45.2 | 32.1 | 43.5 | 7.1 | 16.7 | Pakistan |
| **91** | **601-800** | 4 | Technical University of Madrid | **24.04** | 30.8 | 39.1 | 41.9 | 13.7 | 21.9 | Spain |
| **92** | **601-800** | 5 | Tokyo University of Agriculture and Technology | **23.71** | 20.8 | 44.6 | 20.9 | 22.9 | 26.4 | Japan |
| **93** | **601-800** | 3 | Beijing Institute of Technology | **23.71** | 13 | 78.4 | 19.2 | 24.7 | 30 | China |
| **94** | **601-800** | 4 | Rochester Institute of Technology | **23.45** | 30.9 | 33.7 | 33.8 | 16.2 | 19.8 | USA |
| **95** | **601-800** | 5 | Peter the Great St Petersburg Polytechnic University | **23.40** | 17.1 | 48.8 | 40.1 | 15.9 | 30.9 | Russia |
| **96** | **601-800** | 4 | National University of Sciences and Technology | **22.72** | 36.2 | 32.6 | 35.2 | 8.1 | 19.9 | Pakistan |
| **97** | **601-800** | 4 | AGH University of Science and Technology | **22.59** | 40.3 | 36.6 | 19.4 | 11 | 16.1 | Poland |
| **98** | **601-800** | 2 | Dublin Institute of Technology | **22.18** | 27.7 | 32.1 | 72.6 | 10.4 | 15 | Ireland |
| **99** | **601-800** | 4 | Brno University of Technology | **21.91** | 30.1 | 38.8 | 40.4 | 12 | 17.6 | Czech Republic |
| **100** | **601-800** | 4 | Budapest University of Technology and Economics | **21.76** | 32.2 | 45.8 | 28.5 | 13.6 | 15.8 | Hungary |
| **101** | **601-800** | 3 | Northwestern Polytechnical University | **21.76** | 8.4 | 78.3 | 16.4 | 27.2 | 26.3 | China |
| **102** | **601-800** | 5 | University of Chemistry and Technology, Prague | **21.03** | 23.4 | 39 | 48.6 | 7.2 | 24.1 | Czech Republic |
| **103** | **601-800** | 4 | Birla Institute of Technology and Science, Pilani | **20.98** | 38.9 | 33.1 | 15.5 | 7.5 | 16.9 | India |
| **104** | **601-800** | 5 | University of Science and Technology Beijing | **20.81** | 10.8 | 63.3 | 16.3 | 22.4 | 26.8 | China |
| **105** | **601-800** | 5 | National Institute of Technology Rourkela | **20.41** | 31.7 | 32.7 | 14 | 8.6 | 21.5 | India |
| **106** | **601-800** | 5 | K.N. Toosi University of Technology | **20.18** | 28 | 43 | 14.3 | 13.2 | 18.9 | Iran |
| **107** | **601-800** | 5 | Toyohashi University of Technology | **19.94** | 16.8 | 47.7 | 24.7 | 17.2 | 22.3 | Japan |
| **108** | **601-800** | 5 | Bauman Moscow State Technical University | **19.60** | 3.1 | 44 | 17.4 | 18.3 | 35.9 | Russia |
| **109** | **601-800** | 5 | Nagoya Institute of Technology | **19.57** | 18.5 | 48 | 19.7 | 18.7 | 19.1 | Japan |
| **110** | **601-800** | 5 | Suranaree University of Technology | **19.37** | 22.4 | 33.6 | 31.5 | 10.9 | 20.6 | Thailand |
| **111** | **601-800** | 5 | Izmir Institute of Technology | **18.82** | 25.2 | 42.5 | 29.2 | 9 | 17.7 | Turkey |
| **112** | **601-800** | 5 | TOBB University of Economics and Technology | **18.82** | 28.3 | 32.4 | 30.1 | 10.6 | 13.6 | Turkey |
| **113** | **801+** | 5 | Nagaoka University of Technology | **18.51** | 14 | 38.9 | 31.4 | 12.4 | 24.2 | Japan |
| **114** | **801+** | 5 | National University of Science and Technology (MISiS) | **18.15** | 8.7 | 50.2 | 40.9 | 13.9 | 23.5 | Russia |
| **115** | **801+** | 5 | Gdańsk University of Technology | **17.95** | 28.3 | 37.6 | 17.6 | 9.2 | 14.8 | Poland |
| **116** | **801+** | 3 | King Mongkut's Institute of Technology Ladkrabang | **17.55** | 11.3 | 83.1 | 17.9 | 21.8 | 14 | Thailand |
| **117** | **801+** | 3 | National Taipei University of Technology | **17.50** | 17.9 | 74 | 19.1 | 13.7 | 15.8 | Taiwan |
| **118** | **801+** | 5 | VŠB - Technical University of Ostrava | **17.41** | 20.8 | 37.9 | 27.1 | 11.7 | 15.6 | Czech Republic |
| **119** | **801+** | 5 | Tokyo University of Marine Science and Technology | **17.08** | 11 | 53 | 26.4 | 12.7 | 22.2 | Japan |
| **120** | **801+** | 5 | Wuhan University of Technology | **16.28** | 15.8 | 59.2 | 16.5 | 11.1 | 18.3 | Japan |
| **121** | **801+** | 5 | China University of Mining and Technology | **16.11** | 11.3 | 65.4 | 13.4 | 14.4 | 19.2 | China |
| **122** | **801+** | 5 | Slovak University of Technology in Bratislava | **16.05** | 13.8 | 34.9 | 27.9 | 9.5 | 20.3 | Slovakia |
| **123** | **801+** | 5 | Kyushu Institute of Technology | **15.80** | 14.3 | 50.5 | 22.6 | 10.5 | 18 | Japan |
| **124** | **801+** | 5 | Yıldız Technical University | **15.77** | 22.5 | 44.7 | 18.5 | 8.1 | 13.6 | Turkey |
| **125** | **801+** | 5 | National Research University of Electronic Technology (MIET) | **15.51** | 11.9 | 33.8 | 28.3 | 11.1 | 18.8 | Russia |
| **126** | **801+** | 5 | Riga Technical University | **15.50** | 16.2 | 38.9 | 24.5 | 10.3 | 15.8 | Latvia |
| **127** | **801+** | 5 | Cochin University of Science and Technology | **14.84** | 4.4 | 32.1 | 14.8 | 12.8 | 25.9 | India |
| **128** | **801+** | 3 | Bandung Institute of Technology (ITB) | **13.81** | 1.9 | 84.4 | 28 | 11.1 | 19 | Indonesia |
| **129** | **801+** | 5 | Novosibirsk State Technical University | **13.45** | 3.2 | 37.2 | 31.3 | 10.4 | 20.3 | Russia |
| **130** | **801+** | 5 | University of Electronic Science and Technology of China | **13.37** | 15 | 32.1 | 17.2 | 6.6 | 16 | China |
| **131** | **801+** | 5 | Technical University of Liberec | **12.38** | 3.7 | 35.7 | 28.8 | 9.6 | 17.8 | Czech Republic |
| **132** | **801+** | 5 | Kaunas University of Technology | **12.30** | 4.2 | 36.1 | 21.5 | 10.9 | 17.5 | Lithuania |
| **133** | **801+** | 5 | National Technical University of Ukraine | **12.01** | 1.5 | 33.3 | 17.8 | 9.5 | 21.8 | Ukraine |
| **134** | **801+** | 5 | Chiba Institute of Technology | **11.59** | 8.9 | 34.5 | 15 | 9.6 | 13.5 | Japan |
| **135** | **801+** | 5 | Shibaura Institute of Technology Tokyo | **11.48** | 6.9 | 34.6 | 17.1 | 8.4 | 15.8 | Japan |
| **136** | **801+** | 5 | Vellore Institute of Technology | **11.36** | 7.3 | 33.4 | 20.3 | 7.7 | 15 | India |
| **137** | **801+** | 5 | Lviv Polytechnic National University | **10.25** | 1 | 32.4 | 25 | 7.6 | 16.6 | Ukraine |

**Appendix II. Technical Universities master list (THE Ranking, 2017 edition)**

| DIMENSION | INDICATOR | SCOPE |
|---|---|---|
| Teaching (30%) **Learning environment** | Reputation Survey (15%) | Academic Reputation Survey |
| | Staff-to-student ratio (4.5%) | Data provided by universities |
| | Doctorate-to-bachelor's ratio (2.25%) | Data provided by universities |
| | Doctorates-awarded-to-academic-staff ratio (6%) | Data provided by universities |
| | Institutional income (2.25% | Data provided by universities. Scaled against academic staff numbers and normalised for purchasing-power parity |
| Research (30%) **Volume, income and reputation** | Reputation survey (18%) | Academic Reputation Survey |
| | Research income (6%) | This indicator is fully normalised to take account of each university's distinct subject profile, and it is also normalised for purchasing-power parity |
| | Research productivity (6%) | Number of papers published in the academic journals indexed by Elsevier's Scopus database per scholar, scaled for institutional size and normalised for subject |
| Citations (30%) **Research influence** | | Number of citations made between 2011-2016 to documents published between 2011-2015 according to Scopus |
| International outlook (7.5%) **Staff, students and research** | International-to-domestic-student ratio (2.5%) | Data provided by universities |
| | International-to-domestic-staff ratio (2.5%) | Data provided by universities |
| | International collaboration (2.5%) | Proportion of a university's total research journal publications that have at least one international co-author. Uses the same timespan as Citations indicator. |
| Industrial Income (2.5%) **Knowledge transfer** | | How much research income an institution earns from industry (adjusted for PPP), scaled against the number of academic staff it employs. |

## Appendix III. Technical universities aggregated by cluster and country

| CLUSTER 1 | | CLUSTER 2 | | CLUSTER 3 | | CLUSTER 4 | | CLUSTER 5 | |
|---|---|---|---|---|---|---|---|---|---|
| Uni | Country | Uni | Country | Uni | Country | Uni | Country | Uni | Country |
| 1 | USA | 15 | UK | 20 | China | 39 | USA | 92 | Japan |
| 2 | USA | 22 | Sweden | 28 | Germany | 42 | South Korea | 95 | Russia |
| 3 | UK | 24 | Denmark | 33 | Germany | 44 | Germany | 102 | Czech Republic |
| 4 | Switzerland | 25 | Netherlands | 34 | Japan | 46 | India | 104 | China |
| 5 | Switzerland | 26 | UK | 36 | USA | 48 | Japan | 105 | India |
| 6 | USA | 27 | Germany | 43 | Russia | 51 | Italy | 106 | Iran |
| 7 | Belgium | 29 | Finland | 45 | Germany | 57 | Spain | 107 | Japan |
| 8 | Germany | 30 | France | 52 | Taiwan | 63 | India | 108 | Russia |
| 9 | China | 31 | Australia | 54 | India | 64 | Russia | 109 | Japan |
| 10 | Singapore | 32 | Italy | 56 | India | 65 | Spain | 110 | Thailand |
| 11 | Netherlands | 35 | Sweden | 60 | China | 67 | India | 111 | Turkey |
| 12 | France | 37 | Australia | 61 | India | 73 | China | 112 | Turkey |
| 13 | Germany | 38 | Norway | 68 | Mexico | 76 | Iran | 113 | Japan |
| 14 | Germany | 40 | Austria | 70 | Finland | 77 | Poland | 114 | Russia |
| 16 | South Korea | 41 | Israel | 71 | Iran | 79 | India | 115 | Poland |
| 17 | South Korea | 47 | Australia | 75 | China | 81 | Turkey | 118 | Czech Republic |
| 18 | France | 49 | Italy | 78 | Turkey | 82 | Czech Republic | 119 | Japan |
| 19 | Germany | 50 | Austria | 80 | Iran | 85 | Iran | 120 | Japan |
| 21 | Netherlands | 53 | Cyprus | 83 | China | 86 | USA | 121 | China |
| 23 | Germany | 55 | USA | 88 | Thailand | 87 | Estonia | 122 | Slovakia |
| | | 58 | Saudi Arabia | 89 | China | 90 | Pakistan | 123 | Japan |
| | | 59 | Chile | 93 | China | 91 | Spain | 124 | Turkey |
| | | 62 | UAE | 101 | China | 94 | USA | 125 | Russia |
| | | 66 | USA | 116 | Thailand | 96 | Pakistan | 126 | Latvia |
| | | 69 | New Zealand | 117 | Taiwan | 97 | Poland | 127 | India |
| | | 72 | Finland | 128 | Indonesia | 99 | Czech Republic | 129 | Russia |
| | | 74 | USA | | | 100 | Hungary | 130 | China |
| | | 84 | Jordan | | | 103 | India | 131 | Czech Republic |
| | | 98 | Ireland | | | | | 132 | Lithuania |
| | | | | | | | | 133 | Ukraine |
| | | | | | | | | 134 | Japan |
| | | | | | | | | 135 | Japan |
| | | | | | | | | 136 | India |
| | | | | | | | | 137 | Ukraine |

## TABLES

**Table 1. Summary of the original and simulated weight scores**

| SCENARIO | Research | Citations | Teaching | International outlook | Industry income |
|---|---|---|---|---|---|
| **Original** | 30% | 30% | 30% | 7.5% | 2.5% |
| **Scenario 1 (soft)** | 27.5% | 27.5% | 30% | 7.5% | 7.5% |
| **Scenario 2 (strong)** | 25% | 25% | 30% | 7.5% | 12.5% |
| **Scenario 3 (THE-ENG)** | 30% | 27.5% | 30% | 7.5% | 5% |

**Table 2. Statistical outliers broken down by THE Dimensions**

| Dimension | Tech R |
|---|---|
| Industry income | No outliers |
| Teaching | 1, 2, 3, 4, 5 |
| International outlook | 3, 4, 5, 10, 69 |
| Research | 1, 2, 3, 4, 6, 7, 8, 9, 11 |
| Citations | 1, 2, 3, 5 |

**Table 3. Statistical comparison of Technical Universities and Non Technical Universities**

| Class | Mean values | | | | |
|---|---|---|---|---|---|
| | Industry Income | Teaching | International Outlook | Research | Citations |
| No TU | 44.398 | 29.985 | 48.178 | 25.674 | 50.109 |
| TU | 57.637 | 30.538 | 44.038 | 26.716 | 44.864 |

TU: Technical University

**Table 4. Levene's test (mean) between Technical Universities and Non-Technical Universities**

| Variables | Levene's test | |
|---|---|---|
| | p-value | *Risk (%) |
| Industry Income | **< 0,0001** | **0.01** |
| Teaching | 0.400 | 39.96 |
| International Outlook | **0.021** | <2.07 |
| Research | 0.289 | 28.94 |
| Citations | **0.012** | <1.23 |

* Risk rejecting the null hypothesis H0 (the variances are identical) while it is true

**Table 5. Pearson correlation matrix for THE ranking scores: Technical universities (up) and all universities (bottom)**

| **Scores (n=137)** | Citations | Industry income | International Outlook | Research | Teaching |
|---|---|---|---|---|---|
| Citations | **1** | | | | |
| Industry income | **0.35 | **1** | | | |
| International Outlook | **0.69 | 0.13 | **1** | | |
| Research | **0.74 | **0.57 | **0.57 | **1** | |
| Teaching | **0.71 | **0.53 | **0.54 | **0.94 | **1** |

| **Scores (n= 978)** | Citations | Industry income | International Outlook | Research | Teaching |
|---|---|---|---|---|---|
| Citations | 1 | | | | |
| Industry income | **0.21 | 1 | | | |
| International Outlook | **0.56 | **0.09 | 1 | | |
| Research | **0.65 | **0.43 | **0.43 | 1 | |
| Teaching | **0.58 | **0.38 | **0.29 | **0.90 | 1 |

** Values are different from 0 with a significance level alpha < 0.01

**Table 6. Composition of technical universities clusters**

| Cluster | Size | Average Tech rank | Description |
|---|---|---|---|
| **1** | 20 | 11.0 | World-class TU |
| **2** | 29 | 45.8 | TU focused on international outlook |
| **3** | 26 | 67.8 | National TU focused on industry income |
| **4** | 28 | 73.6 | TU focused on Research impact |
| **5** | 34 | 117.4 | Local TU focused on industry income |

**Table 7. Correlation between dimensions broken down by Cluster**

| Cluster 1 | Industry Income | Teaching | International Outlook | Research | Citations |
|---|---|---|---|---|---|
| Industry Income | **1** | | | | |
| Teaching | -0.27 | **1** | | | |
| International Outlook | -0.53 | 0.33 | **1** | | |
| Research | -0.15 | ****0.85** | 0.38 | **1** | |
| Citations | -0.39 | ****0.73** | 0.37 | ****0.74** | **1** |

| Cluster 2 | Industry Income | Teaching | International Outlook | Research | Citations |
|---|---|---|---|---|---|
| Industry Income | **1** | | | | |
| Teaching | 0.35 | **1** | | | |
| International Outlook | 0.05 | 0.31 | **1** | | |
| Research | 0.10 | ****0.83** | 0.23 | **1** | |
| Citations | -0.03 | ****0.56** | 0.06 | ****0.59** | **1** |

| Cluster 3 | Industry Income | Teaching | International Outlook | Research | Citations |
|---|---|---|---|---|---|
| Industry Income | **1** | | | | |
| Teaching | 0.02 | **1** | | | |
| International Outlook | 0.14 | 0.30 | **1** | | |
| Research | -0.07 | ****0.80** | 0.34 | **1** | |
| Citations | 0.00 | ****0.63** | 0.29 | ****0.55** | **1** |

| Cluster 4 | Industry Income | Teaching | International Outlook | Research | Citations |
|---|---|---|---|---|---|
| Industry Income | **1** | | | | |
| Teaching | 0.36 | **1** | | | |
| International Outlook | -0.31 | -0.03 | **1** | | |
| Research | 0.30 | ****0.81** | -0.01 | **1** | |
| Citations | 0.21 | 0.27 | -0.21 | 0.17 | **1** |

| Cluster 5 | Industry Income | Teaching | International Outlook | Research | Citations |
|---|---|---|---|---|---|
| Industry Income | **1** | | | | |
| Teaching | 0.31 | **1** | | | |
| International Outlook | -0.09 | 0.20 | **1** | | |
| Research | ****0.59** | ****0.64** | -0.11 | **1** | |
| Citations | 0.04 | -0.19 | 0.08 | -0.02 | **1** |

** Values are different from 0 with a significance level alpha < 0.01

## Table 8. Simulation scenarios of Technical universities rankings

| University | Original Score | Soft Score | Strong Score | Tech Score | Original Rank | Soft Rank | Strong Rank | Tech Rank |
|---|---|---|---|---|---|---|---|---|
| California Institute of Technology | 94.3 | 94.0 | 93.6 | 94,1 | 2 | 1 | 1 | 2 |
| Massachusetts Institute of Technology | 93.4 | 93.0 | 92.6 | 93,1 | 5 | 3 | 2 | 3 |
| Imperial College London | 90.0 | 88.8 | 87.6 | 89,3 | 8 | 7 | 6 | 7 |
| ETH Zurich | 89.3 | 87.8 | 86.3 | 88,5 | 9 | 9 | 9 | 9 |
| École Polytechnique Fédérale de Lausanne | 76.8 | 76.2 | 75.6 | 76,1 | 30 | 32 | 30 | 32 |
| Georgia Institute of Technology | 76.3 | 75.1 | 74.0 | 75,5 | 33 | 34 | 35 | 34 |
| KU Leuven | 73.8 | 74.7 | 75.6 | 74,0 | 40 | 35 | 31 | 39 |
| Technical University of Munich | 71.5 | 72.7 | 73.9 | 72,0 | 46 | 41 | 36 | 45 |
| Hong Kong University of Science and Technology | 71.1 | 70.2 | 69.4 | 70,4 | 49 | 50 | 51 | 50 |
| Nanyang Technological University | 70.0 | 70.9 | 71.8 | 70,0 | 54 | 49 | 43 | 51 |
| Delft University of Technology | 67.9 | 69.4 | 70.8 | 68,7 | 59 | 53 | 48 | 56 |
| École Normale Supérieure | 65.8 | 64.3 | 62.9 | 64,6 | 66 | 69 | 79 | 67 |
| RWTH Aachen University | 63.0 | 64.6 | 66.2 | 63,7 | 78 | 67 | 61 | 72 |
| Technical University of Berlin | 62.0 | 63.6 | 65.2 | 62,6 | 82 | 73 | 64 | 78 |
| University of Warwick | 62.0 | 60.7 | 59.4 | 61,0 | 82 | 90 | 103 | 87 |
| Korea Advanced Institute of Science and Technology | 61.3 | 63.0 | 64.7 | 61,8 | 89 | 76 | 68 | 82 |
| Pohang University of Science and Technology | 59.6 | 61.3 | 63.1 | 60,1 | 104 | 88 | 73 | 95 |
| École Polytechnique | 58.6 | 59.5 | 60.4 | 58,8 | 116 | 103 | 92 | 107 |
| Karlsruhe Institute of Technology | 55.8 | 57.7 | 59.5 | 55,8 | 144 | 116 | 99 | 134 |
| University of Science and Technology of China | 54.7 | 55.7 | 56.6 | 54,7 | 153 | 144 | 126 | 148 |

**Table 9. Rank position of Technical universities for each simulated scenario**

| Scenarios | Top 50 | Top 100 | Top 200 | Top 300 | Top 400 | Top 500 | Top 600 | Top 700 | Top 800 | Top 900 | All |
|---|---|---|---|---|---|---|---|---|---|---|---|
| Original | 9 | 16 | 27 | 40 | 41 | 60 | 78 | 94 | 111 | 126 | 137 |
| Scenario 1 (soft) | 10 | 17 | 28 | 41 | 52 | 63 | 80 | 95 | 114 | 127 | |
| Scenario 2 (strong) | 10 | 18 | 28 | 42 | 53 | 68 | 83 | 95 | 115 | 126 | |
| Scenario 3 (THE-ENG) | 9 | 17 | 27 | 41 | 51 | 61 | 80 | 96 | 114 | 128 | |

**FIGURES**

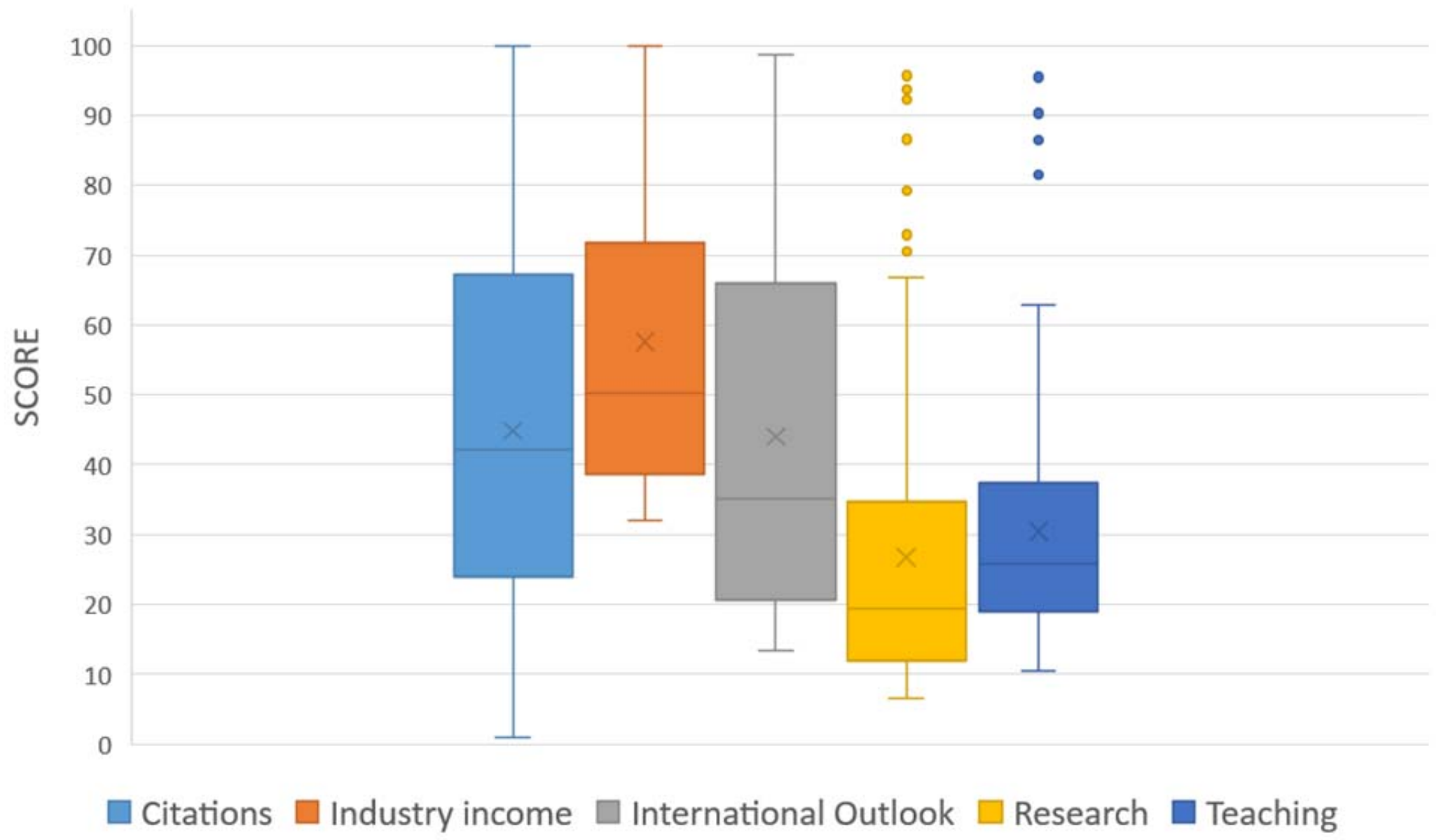

**Figure 1. Statistical performance of dimension scores for Technical Universities**

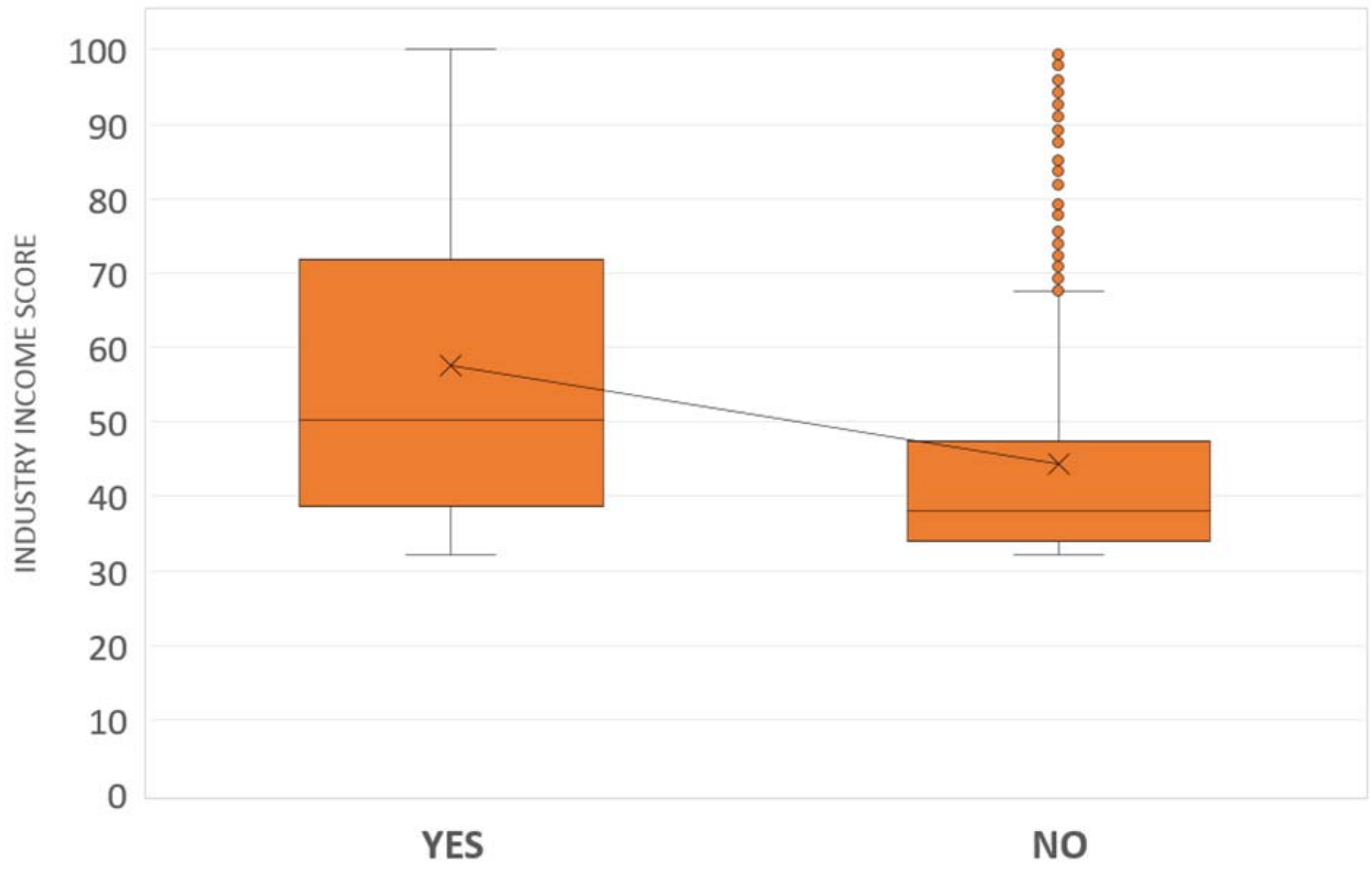

**Figure 2. Industrial income for Technical and non-Technical Universities**

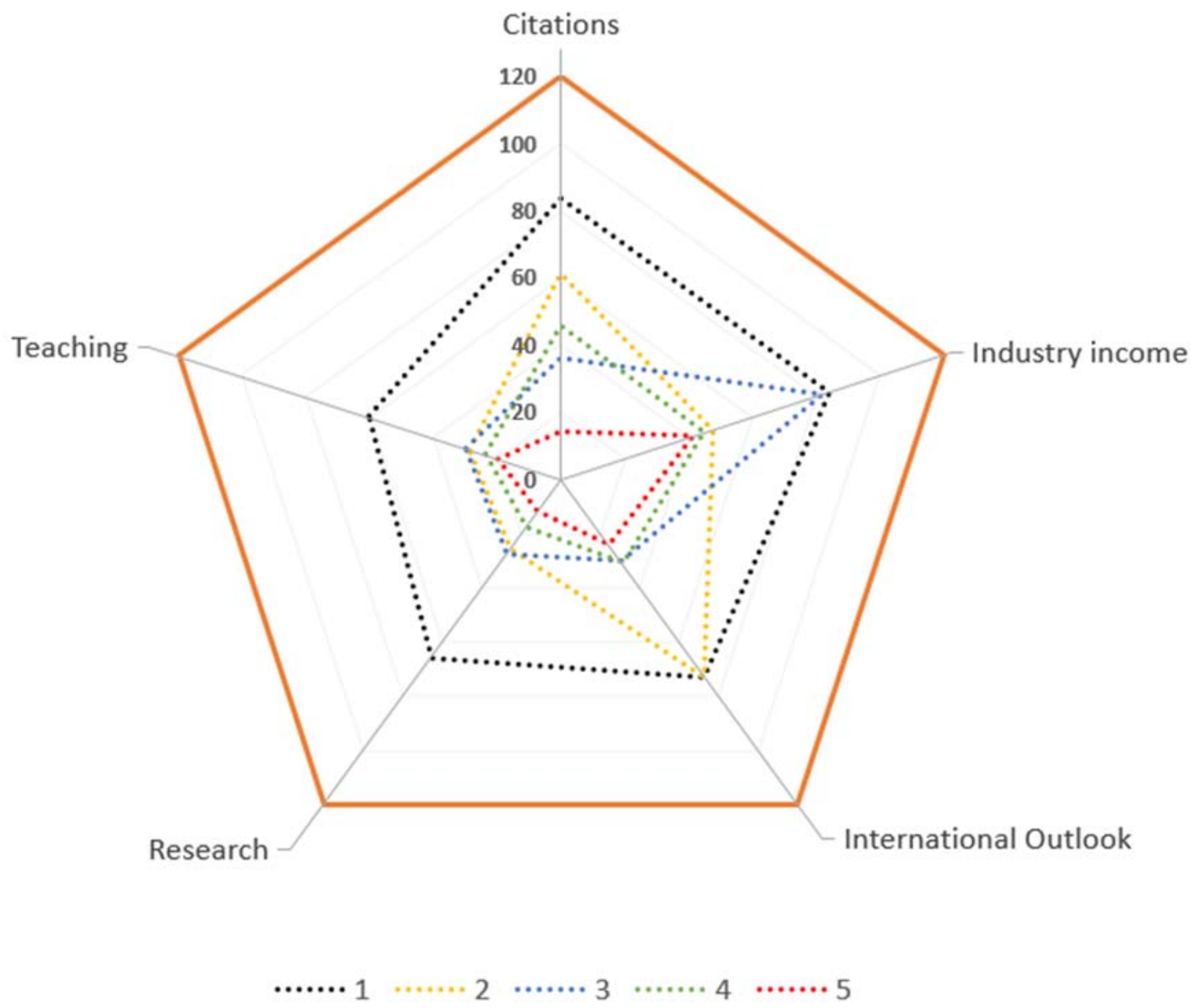

**Figure 3. Technical Universities clusters according to THES ranking scores**

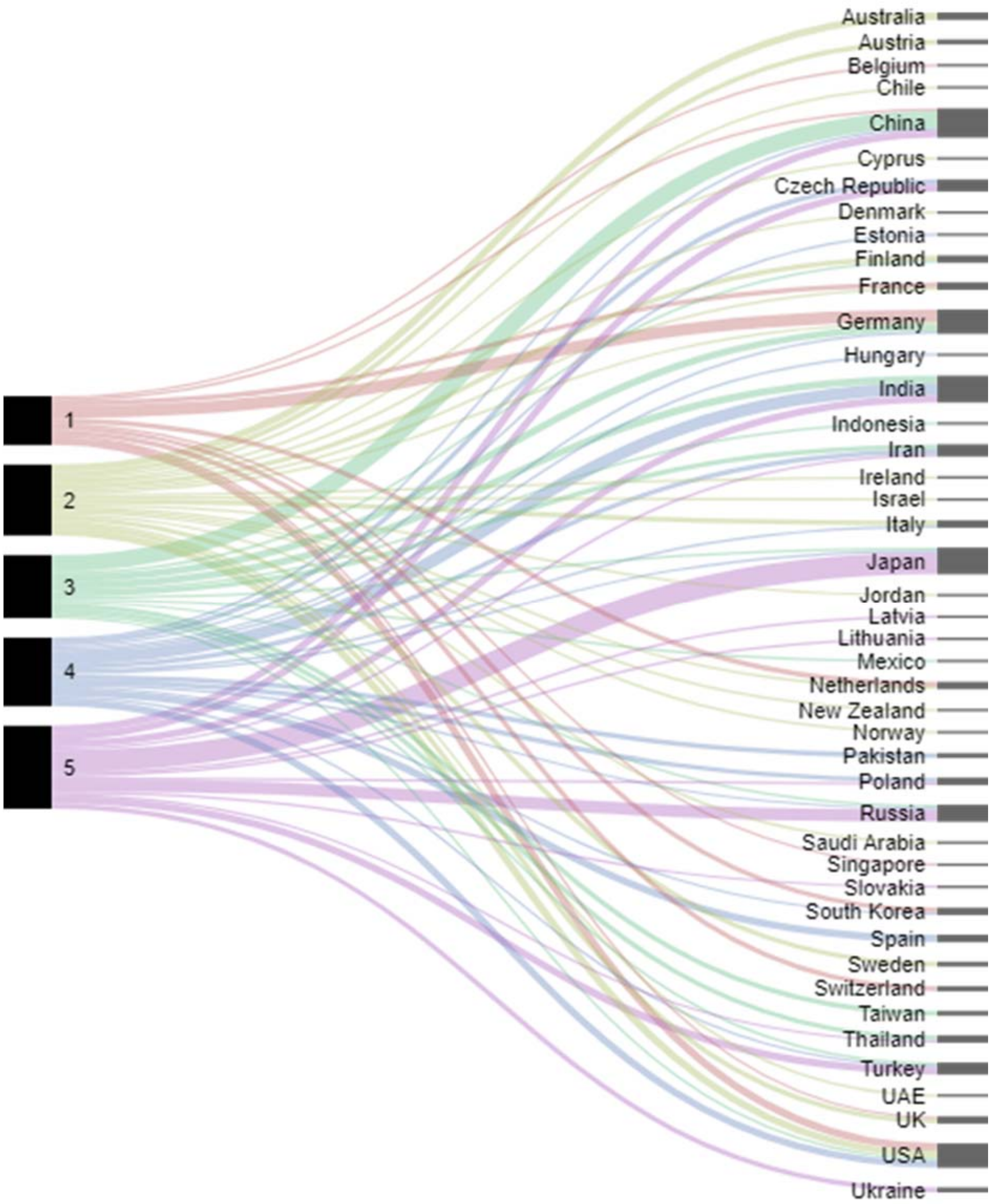

**Figure 4. Distribution of countries broken down by cluster**

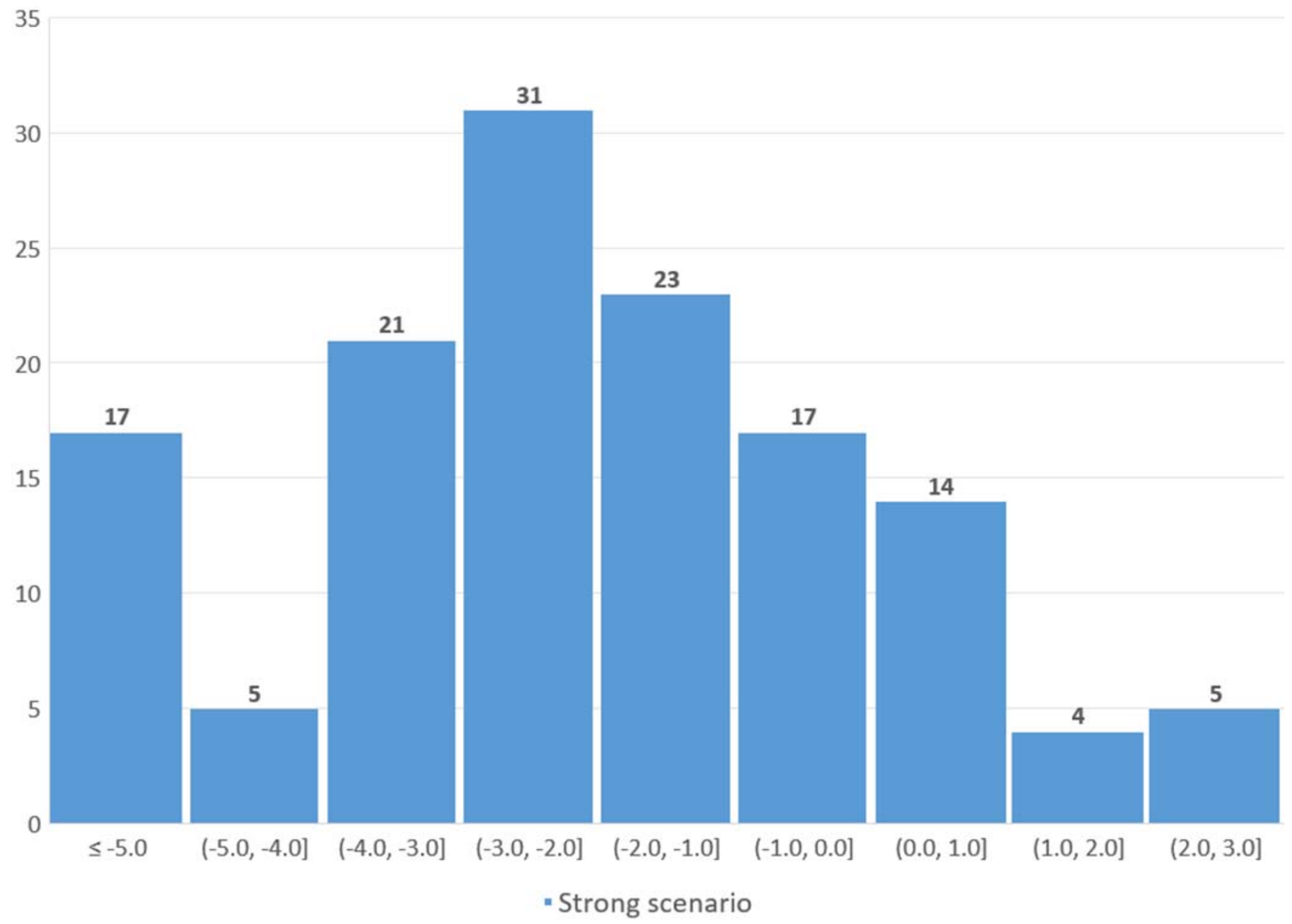

**Figure 5. Histogram of difference between original and strong scenario scores for Technical universities**